\documentclass[twocolumn,aps,prl,final,groupedaddress]{revtex4}

\usepackage{amssymb}

\usepackage[pdftex]{graphicx}

\renewcommand{\deg}{^\circ}

\begin{document}

\title{Phonon Softening in Superconducting Diamond}

\author{M.~Hoesch}
\altaffiliation[current affiliation ]{ESRF, 6 rue Jules Horowitz, 38000 Grenoble, France}
\email{hoesch@esrf.fr}
\affiliation{SPring-8, JAEA, 1-1-1 Kouto, Sayo, Hyogo, Japan}

\author{T.~Fukuda}
\affiliation{SPring-8, JAEA, 1-1-1 Kouto, Sayo, Hyogo, Japan}

\author{T.~Takenouchi}
\affiliation{School of Science and Engineering, Waseda University, 3-4-1 Okubo, Shinjuku, Tokyo, Japan}

\author{J.P.~Sutter}
\altaffiliation[new affiliation ]{Diamond Light Source, Chilton, Didcot, Oxfordshire, England}
\affiliation{SPring-8, JASRI, 1-1-1 Kouto, Sayo, Hyogo, Japan}

\author{S.~Tsutsui}
\affiliation{SPring-8, JASRI, 1-1-1 Kouto, Sayo, Hyogo, Japan}

\author{A.Q.R.~Baron}
\affiliation{RIKEN, 1-1-1 Kouto, Sayo, Hyogo, Japan}
\affiliation{SPring-8, JASRI, 1-1-1 Kouto, Sayo, Hyogo, Japan}

\author{M.~Nagao}
\affiliation{National Institute for Materials Science, 1-2-1 Sengen, Tsukuba, Japan}

\author{Y.~Takano}
\affiliation{National Institute for Materials Science, 1-2-1 Sengen, Tsukuba, Japan}

\author{H.~Kawarada}
\affiliation{School of Science and Engineering, Waseda University, 3-4-1 Okubo, Shinjuku, Tokyo, Japan}

\author{J.~Mizuki}
\affiliation{SPring-8, JAEA, 1-1-1 Kouto, Sayo, Hyogo, Japan}

\date{\today}

\begin{abstract}

We observe strong softening of optical phonon modes in superconducting ($T_c=4.2$~K) boron-doped diamond near the Brillouin zone center using inelastic x-ray scattering from a CVD-grown highly oriented sample.  
The magnitude of the softening, and our observation that it becomes
stronger approaching zone center, supports theoretical models
suggesting  a phonon-mediated pairing mechanism via coupling of
optical  phonon modes to Fermi surfaces around the zone center. The
electron-phonon coupling parameter is determined as approximately $\lambda = 0.33$.

\end{abstract}


\maketitle

Pure diamond is a wide bandgap insulator. When it is boron-doped
beyond  the metal-to-insulator transition  at $n_B \approx 3\cdot 10^{20}
\ \mathrm{cm}^{-3}$, it shows superconductivity with transition
temperatures  $T_c$ that are remarkably high for an impurity doped
material  with low carrier density. Superconductivity was reported
both in  samples grown by high-pressure-high-temperature synthesis
with  $T_c=4$~K at a B-doping level $n_B \approx 5 \cdot 10 ^{21}
\mathrm{cm}^{-3}$~\cite{ekimov04} and in samples grown by $\mu$-wave assisted
chemical vapor deposition (CVD) \cite{takano04,bustarret04}. $T_c$
depends on the doping level $n_B$ and increases up to $T_c^{on} \approx 11$~K (onset) 
at the highest doping reported so far~\cite{umezawa05}.

Conventional theory of phonon-mediated superconductivity suggests that
an ultrahard material like diamond can have a high $T_c$ if the high frequency modes
have  a strong enough electron-phonon coupling (EPC). The
bond-stretching  optical modes of diamond around $\omega_0 = 160$~meV naturally couple to the
covalent  bonding states of the diamond valence band.
Various ab-initio models have predicted the EPC coupling constant
$\lambda$.  In the typical doping range between $n = 2.5 - 3$~at\%, 
values between between $\lambda=0.27$ (virtual crystal
approximation VCA~\cite{boeri04}) and $\lambda=0.39$ (supercell~\cite{xiang04}) 
are calculated and lead to predicted $T_c= 0.2 - 4.4$~K that correspond well 
to those observed in real samples. 
The nature of the modes that couple to the
free charge carriers is, however, difficult to assess by theory 
and few experimental data are available. VCA naturally focuses on the 
optical modes of the diamond lattice that couple due to their
bond-stretching  nature~\cite{lee04,boeri04}. The $\lambda$ calculated
in a supercell also includes localized B-modes, where the coupling is
enhanced due to the enhanced valence band wave functions
at the dopant sites~\cite{blase04}. 
 
We have studied the lattice dynamics in a sample of highly B-doped superconducting
diamond. For the optical phonon branches, we observe a 
softening that allows the determination of
the momentum-dependent EPC parameter $\lambda(q)$. 
Similar to inelastic neutron scattering, the technique of inelastic x-ray scattering (IXS) probes the
momentum-dependent frequency of the phonon excitations. A strong
softening of the optical phonon branches is the fingerprint of a
strong EPC. Moreover, the observation of the
momentum-dependence of the softening provides information on the
phase-space of the coupling. This complements early indications of
phonon softening
from Raman scattering experiments that probe only zero momentum~\cite{okano90}. 
In the case of high B-doping, Raman scattering
shows a strong signal from both lattice as well as electronic
excitations that could not be disentangled~\cite{bustarret03} and
thus no quantitative data on the EPC have been gained from the Raman
spectra. A recent study of far-infrared and THz-radiation absorption spectra
shows one peak of EPC in the region of the optical phonons as
well as contributions in a region of lower frequency modes~\cite{ortolani06}.

The electronic structure in the metallic regime is
described either by a shift of the Fermi level into the diamond
valence band (degenerate metal) or by a purely B-related impurity band or a mixture of
both. Experimentally, a Fermi
level in the diamond valence band was observed by angle-resolved
photoemission spectroscopy (ARPES)~\cite{yokoya05}, but evidence for a persistence
of an impurity band close to the top of the valence band maximum was
reported by optical spectroscopy~\cite{wu05}. The former scenario
would lead to a well-defined ``spheroid'' Fermi surface (FS) around the
$\Gamma$-point. This FS has significant anisotropy due
to the covalent bonding states and consists of three concentric sheets. The
latter scenario should not lead to a well-defined FS if the
dopant atoms occupy random positions in the lattice. Calculations using
the coherent phase approximation suggest the formation of a spheroid
FS around $\Gamma$ but a smearing of the Fermi
wave vector due to impurity scattering~\cite{lee05}.

In the presence of a spheroid FS a strong phonon softening is expected
at small momentum $q$, where energy-conserving phonon scattering is
possible within the FS. The softening is
gradually reduced to a constant value at increased phonon momentum
around  the Brillouin zone boundary.  
This gentle kind of a Kohn anomaly is reminiscent of the EPC in 
MgB$_2$, where the Cooper pairs form due to a
coupling of the electronic $\sigma$-band to the E$_\mathrm{2g}$
optical phonons~\cite{boeri04,shukla03,baron04}. 

\begin{figure}
\centerline{\includegraphics[width = .34\textwidth]{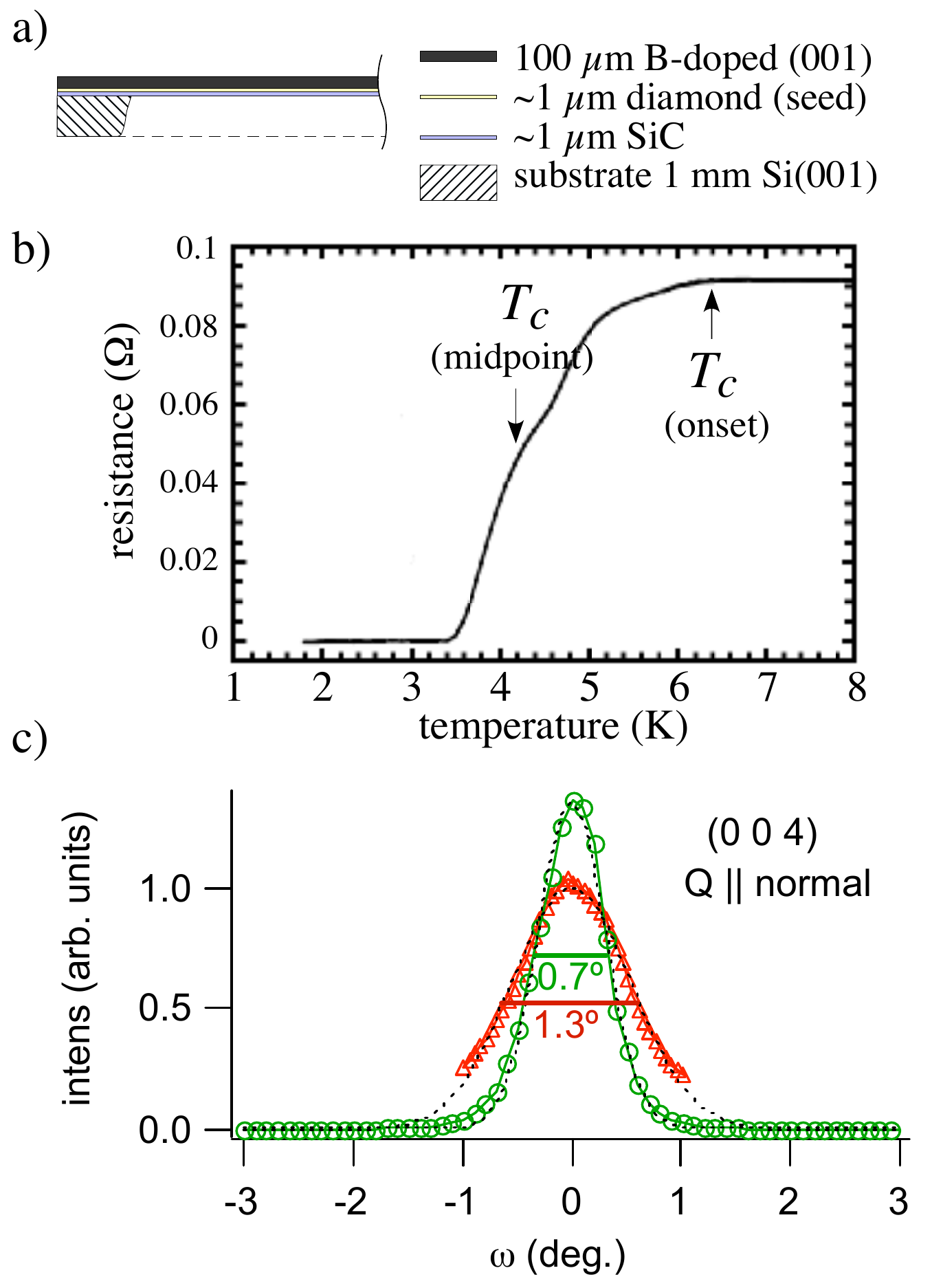}}
\caption{(a) Schematic view of the $\sim\!\!100 \mu\mathrm{m}$
  thick  B-doped sample B. (b) Temperature-dependent resistance curve.
  Above $T_c$ the resistance is almost flat with a room temperature
  resistivity of $\sim\!\!15\ \Omega \mathrm{cm}$. (c) X-ray
  rocking curves of the (004) reflection of sample B ($\circ$) and
  undoped sample N ($\vartriangle$).}
\label{samplecharact}
\end{figure}

The disorder in B-doped diamond has only  a little direct effect
on the lattice dynamics, because the mass difference of B- and C-atoms
is small. Like the natural isotope mixture of C-atoms in diamond it
would have a significant effect only at doping levels much higher than a
few atomic percent~\cite{widulle02}. In our data we
found evidence for a slight disorder-induced change of the Eigenvectors of
acoustic modes~\cite{hoesch06}, but no evidence for an
EPC of acoustic or localized modes was found. In this letter we
discuss the coupling of the optical phonons, where a direct comparison
between pure and doped diamond leads to a quantitative determination
of the mode-dependent coupling parameter $\lambda$.

The samples were grown by heteroepitaxial CVD deposition on a Si(001)
substrate pre-treated by SiC to yield a highly ordered (001)-oriented
ensemble of coalesced diamond grains~\cite{kawarada97}. A layer of 100 $\mu$m B-doped diamond
at a doping level of $n_B = 3.8\cdot10^{21}\ \mathrm{cm}^{-3} \approx 2$~at\%
was deposited in $\sim$100 hours (sample B). The grains are 
oriented within 1$\deg$ as seen in the x-ray rocking curve in 
Fig.~\ref{samplecharact}c. The superconducting transition is rather broad
 around $T_c = 4.2$~K (see Fig.~\ref{samplecharact}b). 
The silicon substrate was removed by chemical
etching. A second sample of  100 $\mu$m of more lightly nitrogen-doped diamond 
($n_N = 4 \cdot10^{18}\ \mathrm{cm}^{-3}$) was grown by the same
technique (sample N). Nitrogen is introduced to
enhance the CVD growth. It does not lead to an electronic doping due
to its deep impurity level. Our experiment confirmed the phonon
dispersion of sample N to be identical to
pure diamond~\cite{kulda02} and we will refer to this reference sample
as undoped. The doping levels were determined by secondary ion mass
spectroscopy (SIMS).

The experiment was performed at the IXS station
of BL35XU~\cite{baron00} at the SPring-8 synchrotron light source. 
X-rays of $h\nu =15.82$~keV were selected with energy resolution of
6.4~meV by use of the Si(888) reflection in the backscattering
monochromator and in the diced spherical analyzers. For each of the
two samples 8 positions of the diffractometer arm were measured, which
results in 8 spectra precisely on the high symmetry lines $\Gamma$-L
and $\Gamma$-X. Additional 21 spectra very close to the high symmetry
line at identical positions for undoped and B-doped diamond were also
analyzed. These spectra show only longitudinal phonons (LO) due to the
measurement geometry. In addition, a spectrum was
acquired at the L-point for the transverse phonons (TA and TO). 
The energy scale was calibrated  for each analyzer using the elastic
line position and the
optical phonon peak from a pure single crystal diamond sample close
to the $\Gamma$-point at $Q=(1.1\ 1.1\ 1.1)$~\cite{kulda02}. All data
were acquired at room temperature because the EPC
as such has very weak temperature dependence and can be studied much
more conveniently at room temperature than at lower temperature. We
remark at this point that even at temperatures below $T_c$ no change of
the EPC is expected for the optical modes that
have frequencies $\omega_0$ much larger than the superconducting gap
$\omega_0 \gg 2\Delta \approx 1$~meV~\cite{ishizaka06,ortolani06}.

\begin{figure*}
\centerline{\includegraphics[width = 0.78\textwidth]{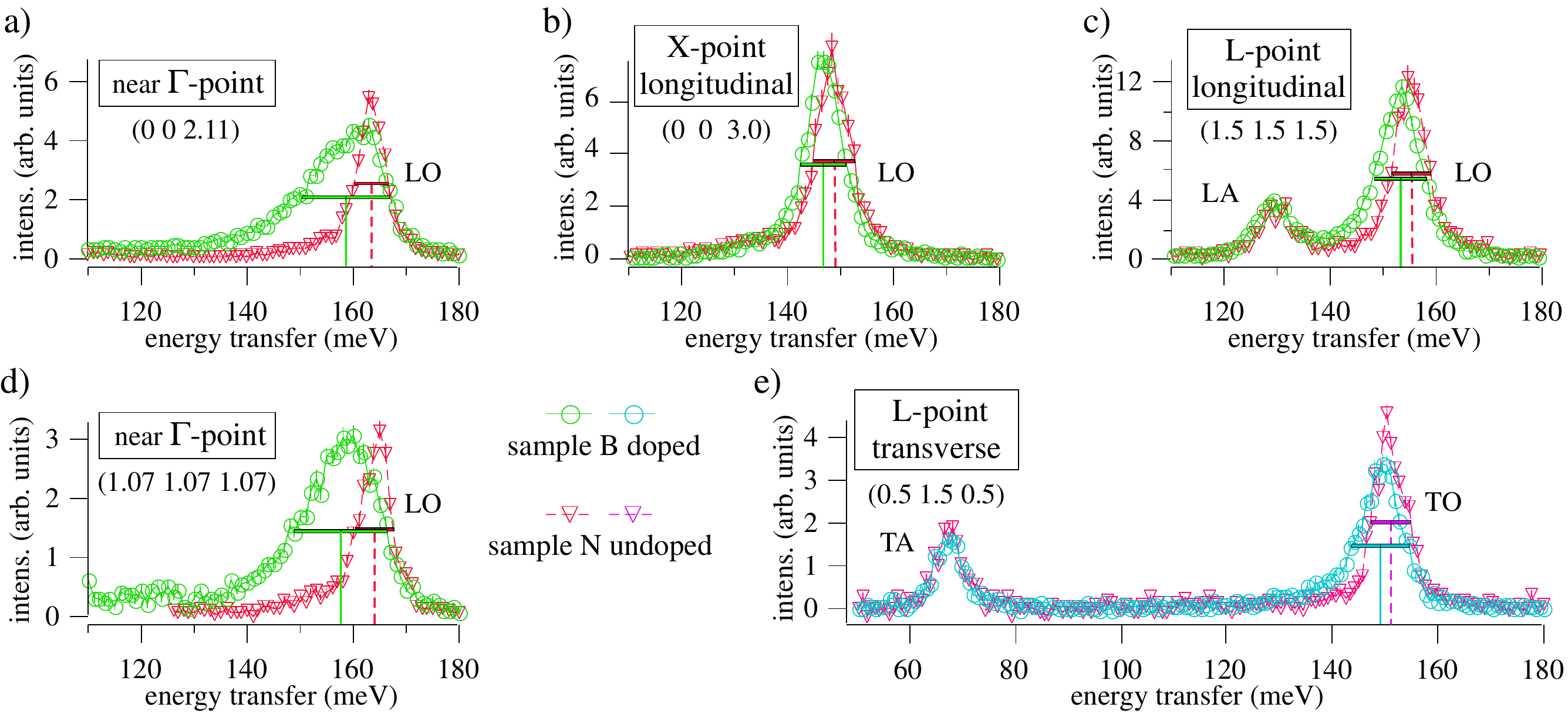}}
\caption{IXS spectra from undoped sample N ($\triangledown$) and sample B
  ($\circ$) for momentum transfer along the [001]-direction $\Gamma$-X (a and b)
  and the [111] direction  $\Gamma$-L (c and d) in a geometry where only
  longitudinal phonons contribute, as well as at $Q= (0.5\ 1.5\ 0.5)$
  (e) where mostly transverse phonons contribute. The total momentum
  transfer $Q$ is indicated for each pair of spectra in relative
  lattice units. The peak position and width at half height is indicated.}
\label{ixs_spectra}
\end{figure*}

\begin{figure}[b!]
\centerline{\includegraphics[width = .45\textwidth]{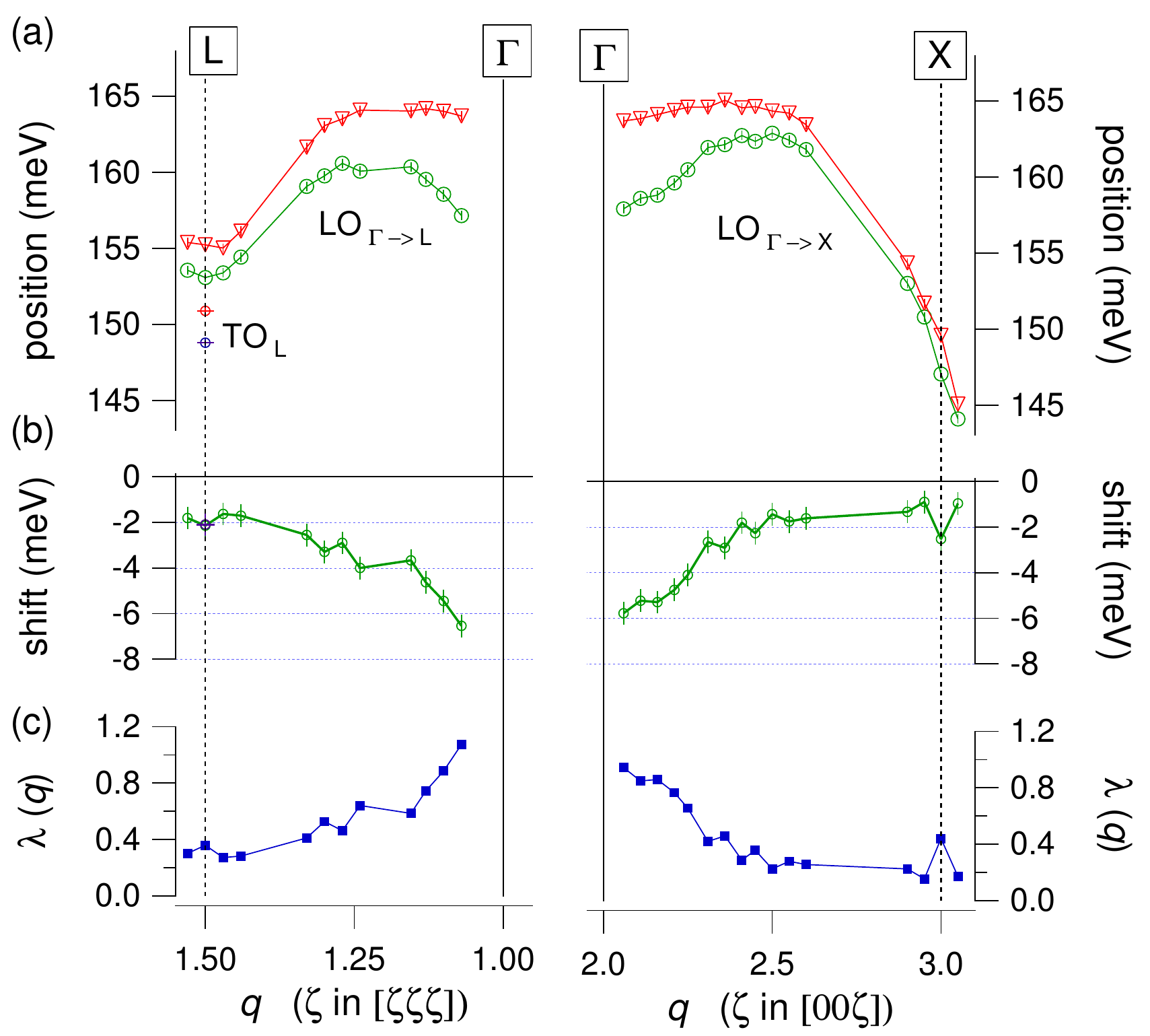}}
\caption{(a) Dispersion of the optical phonons in sample
  N~($\triangledown$) and B~($\circ$) as determined by the peak-positions 
  as shown in Fig.~\ref{ixs_spectra}. In the
  left-hand side  the $\Gamma$-L and in the right-hand side
  the $\Gamma$-X direction is shown. Most data are from the
  LO  phonons, but in the left-hand side at L the TO phonons are also
  indicated. (b) Softening curve: peak shift for sample B with respect
  to sample N as a function of momentum. (c) Momentum-dependent
  EPC parameter $\lambda(q)$ as determined by
  Eq.~(\ref{EqlambdaQ}).}
\label{dispersion}
\end{figure}

A selection of measured spectra in the region of the optical phonons
is shown in Fig.~\ref{ixs_spectra}. For small momenta close to the
$\Gamma$-point, the peaks are strongly shifted with respect to undoped
diamond (extrapolated $-8$~meV). At the Brillouin zone boundary the shift is smaller  
($-2$~meV) both for the X-point and the L-point and for LO as well as for TO
phonons. At the zone boundary the peak shape is given by the
resolution function of the IXS spectrometer. Near the zone center the
peaks from sample N are still resolution-limited, while the peaks from
sample B are very broad and asymmetric. This asymmetric peak shape is
incompatible with a life-time reduction of the phonons in a damped
harmonic oscillator theory. More probably it is given by a certain
degree of in-homogeneity of the B-content across the sample that is
also evident in the broad superconducting transition shown in 
Fig.~\ref{samplecharact}b. The same peak positions and asymmetric shapes 
were recently also observed in a sample grown by homo-epitaxial 
CVD~\cite{hoesch06b}.

The dispersion shown in Fig.~\ref{dispersion}a is
determined from the peaks as the arithmetic middle between the two
flanks at half height as indicated in Fig.~\ref{ixs_spectra}. The
difference of the dispersions for samples N and B is shown in
Fig~\ref{dispersion}b. Towards $\Gamma$, the peak shift approaches an
extrapolated value of 8~meV. This value is larger than the one naively derived
from the shift in Raman spectra ($<4$~meV), where the one-phonon line remains
barely visible above the background of the lower-energy Fano-feature. 
The momentum-dependence of the softening is compatible with a gentle Kohn Anomaly in the spheroid FS of a 
degenerate metal, as the two directions and also longitudinal and
transverse phonons soften in the same way. Thus, we find that the phonon
softening is purely dependent on the absolute value of the momentum
$q$ for the two high symmetry direction we investigated.

To determine the $q$-dependent EPC parameter 
$\lambda(q)$, the first
approach would be to consider the life-time broadening of the phonons
and use Allen's formula that connects $\lambda(q)$ to the peak width
$\gamma(q)$. In diamond, this approach fails because of the
in-homogeneity induced broadening of the LO peaks near $\Gamma$ and
their asymmetric line-shape. Instead, we can determine $\lambda(q)$
from the softening from Allen's analysis \cite{allen72} as
\begin{equation}
\lambda(q) = \frac{1}{N(0)E_c}\left(\frac{\omega^2_N(q)}{\omega^2_B(q)} - 1 \right),
\label{EqlambdaQ}
\end{equation}
where $E_c$ is a cut-off energy identified with the Fermi energy
$E_c=E_F$, $N(0)$ is the density of states at $E_F$
(in states/eV/cell) and $\omega(q)$ is the measured dispersion for
samples N and B. We take  $N(0)$ and $E_F$ from VCA calculations for our
doping level ($N(0) = 0.12$~states/eV/cell and $E_F =
0.62$~eV)~\cite{lee04}. The product $N(0) E_F$ is proportional to the
number of holes per unit cell
(in a parabolic band in diamond $N(0) E_F = 3n$), so that different models 
of the band dispersion still give only a small error in the determination 
of $\lambda(q)$. The relevant EPC parameter
is the average over the Brillouin zone $\lambda=
1/V_{BZ}\int_{BZ}\lambda(q) d^3q$. Assuming isotropic softening we can
extrapolate $\lambda(q)$ from either of the two measured momentum
directions independently and determine the Brillouin zone average as 
$\lambda_{\Gamma \mathrm{L}} =  0.39$ and $\lambda_{\Gamma \mathrm{X}} =
0.27$, respectively. The difference gives an approximate measure of
the experimental uncertainty of
this analysis. The final value is the average between the two: $\lambda
= 0.33 \pm 0.06$. The systematic uncertainty due to the theoretical
approximations is probably of the same magnitude ($\pm 20 \%$).

The experimental EPC parmeter lies in the range of theoretical
predictions. When inserted into the McMillan relation $T_c = (\omega_0/1.2)\exp\left[-1/\left(\lambda/(1+\lambda)
-\mu^*\right)\right]$ using the
optical phonon frequency $\omega_0 = 160$~meV and a conventional $\mu^*=0.12$,
a $T_c = 0.63$~K is derived ($T_c=3$~K at the higher error limit of
$\lambda$), which is significantly smaller than the observed $T_c = 4.2$~K. 
In a direct comparison the measured softening is less strong than the predictions
by VCA calculations ($\sim$30 meV at $\Gamma$ and  $\sim$6~meV at the
L-point~\cite{boeri06}). Supercell calculations that provide a more realistic
incorporation of the B-atoms into the diamond lattice suffer from
ordered-boron artifacts, but the softening reported a a higher doping
level of 6.25~at\% (12.5 meV at $\Gamma$~\cite{xiang04}) is in better agreement with 
our observations. In addition to the optical phonons, these
calculations also include coupling to localized B-modes. 
Thus, although the observed coupling of the optical phonons is strong,
there is room for other modes to provide additional EPC. 

From our
data the determination of $\lambda$ is only possible for those modes that
have a direct relation between the doped and the undoped case, i.e. the
six phonon branches of the diamond lattice. No coupling was observed
for the acoustic modes~\cite{hoesch06}. Also, we have not observed
a clear signature from B-derived, localized modes. In the IXS
spectra we have, however, observed finite intensity almost throughout
the spectral range that is probably related to impurity modes. Our
data are thus consistent with a recent optical spectroscopy experiment
that found significant contributions to the EPC
both in the high-frequency spectral range above
$\omega = 140$~meV as well as in the lower range around $\omega =
70$~meV~\cite{ortolani06}. The former region corresponds well to the
optical phonon coupling reported in this letter. 
The latter region is also the range of B-B
pair vibrations that lead to a very large Raman scattering
signal but do not contribute to the EPC~\cite{bourgeois06}. 
Contributions from single boron localized modes are also predicted around
70~meV~\cite{xiang04,bourgeois06}, matching the analysis of
Ref.~\cite{ortolani06}, but so far unaccounted by direct spectroscopy.

In conclusion, we have measured the softening of the optical phonons
in diamond. We found a strong softening with a characteristic momentum
dependence both along $\Gamma$-L and $\Gamma$-X. This momentum
dependence is compatible with coupling to the holes in a spheroid
FS as predicted by theory for a Fermi level in the valence
band of diamond. We have determined the momentum dependent
EPC parameter $\lambda(q)$ for the optical phonons
and its average over the Brillouin zone is the mode-specific coupling parameter $\lambda =
0.33\pm 0.06$. Our data thus strongly support an Eliashberg model of
the superconductivity in B-doped diamond with coupling through the
bond-stretching optical modes that are particularly high in frequency
for the ultrahard diamond material.

We would like to thank the Japan Synchrotron Radiation Research
Institute (JASRI) for granting beamtime under proposal numbers
2004B0736 and 2005A0596. Dr. Sakaguchi at NIMS
characterized the samples by SIMS. Fruitful discussion with
M.~Tachiki, E.~Bustarret, L.~Boeri and A.~Mirone is gratefully
acknowledged. This work was supported by the Grant-in-Aid for
Scientific Research on Priority Areas "Invention of anomalous quantum
materials" from the MEXT. One of us (MH) would like to thank the
Japanese Society for the Promotion of Science (JSPS) for financial support.

\end{document}